\documentclass[12pt,preprint]{article}

\usepackage{amssymb}
\usepackage{amsmath}
\usepackage{graphics}
\usepackage{epsfig}
\renewcommand{\vec}[1]{{\bf #1}}
\newcommand{\eqb}{\begin{equation}}
\newcommand{\eqe}{\end{equation}}
\newcommand{\dmb}{\begin{displaymath}}
\newcommand{\dme}{\end{displaymath}}
\newcommand{\pd}{\partial}

\newcommand{\eab}{\begin{eqnarray}}
\newcommand{\eae}{\end{eqnarray}}

\newcommand{\e}{\mbox{e}}
\newcommand{\be}{\begin{equation}}
\newcommand{\ee}{\end{equation}}

\begin{document}
\begin{titlepage}

\begin{center}
\Large{SU(2) Yang-Mills Theory:\\ Waves, Particles, and Quantum Thermodynamics}\vspace{0.3cm}\\ 
\large{Ralf Hofmann}
\end{center}
\vspace{0.1cm} 
\begin{center}
{\em 
Institute for Photon Science and Synchrotron Radiation\\ 
Karlsruhe Institute of Technology\\  
Hermann-von-Helmholtz-Platz 1\\ 
76344 Eggenstein-Leopoldshafen, Germany}
\vspace{0.5cm}\\ 
{\em  Institut f\"ur Theoretische Physik\\ 
Universit\"at Heidelberg\\ 
Philosophenweg 16\\ 
69120 Heidelberg, Germany}
\end{center}
\vspace{.5cm}
\begin{abstract}
 We elucidate how Quantum Thermodynamics at temperature $T$ emerges from 
pure and classical SU(2) Yang-Mills theory on a four-dimensional Euclidean spacetime 
slice $S_1\times {\bf R}^3$. The concept of a (deconfining) thermal 
ground state, composed of certain solutions to the fundamental, classical Yang-Mills equation, 
allows for a unified addressation of both (classical) wave- and 
(quantum) particle-like excitations thereof. More definitely, the thermal ground state represents the interplay between nonpropagating, periodic configurations which are electric-magnetically (anti)selfdual in a non-trivial way and possess topological charge modulus unity. Their trivial-holonomy versions --  Harrington-Shepard (HS) (anti)calorons -- yield an accurate {\sl a priori} estimate of 
the thermal ground state in terms of spatially coarse-grained centers, each containing one 
quantum of action $\hbar$ localized at its inmost spacetime point, which induce an inert adjoint 
scalar field $\phi$ ($|\phi|$ spatio-temporally constant). 
The field $\phi$, in turn, implies an effective pure-gauge configuration, $a^{\rm gs}_\mu$, 
accurately describing HS (anti)caloron overlap. Spatial homogeneity of the thermal 
ground-state estimate $\phi,a^{\rm gs}_\mu$ demands that (anti)caloron centers 
are densely packed, representing a collective departure from (anti)selfduality. 
Effectively, such a ``nervous'' microscopic situation gives rise to two 
static phenomena: finite ground-state energy density $\rho_{\rm gs}$ and pressure 
$P_{\rm gs}$ with $\rho_{\rm gs}=-P_{\rm gs}$ as well as the (adjoint) Higgs mechanism. 
The peripheries of HS (anti)calorons are static and resemble (anti)selfdual 
dipole fields whose apparent dipole moments are determined by $|\phi|$ and $T$, protecting them 
against deformation potentially caused by overlap. Such a protection extends to the spatial density of 
HS (anti)caloron centers. Thus the vacuum electric permittivity $\epsilon_0$ and magnetic 
permeability $\mu_0$, supporting the propagation of wave-like disturbances 
in the u(1) Cartan subalgebra of su(2), can be reliably calculated for disturbances which 
do not probe HS (anti)caloron centers. Both $\epsilon_0$ and $\mu_0$ turn out to be 
temperature independent in thermal equilibrium but also for an 
isolated, monochromatic u(1) wave. HS (anti)caloron centers, on the other hand,  
react onto wave-like disturbances, which would resolve their spatio-temporal structure, by 
indeterministic emissions of quanta of energy and momentum. 
Thermodynamically seen, such events are Boltzmann weighted and occur 
independently at distinct locations in space and 
instants in (Minkowskian) time, entailing the Bose-Einstein distribution. 
Small correlative ramifications associate with effective radiative corrections, e.g., in terms of polarization tensors. We comment on an SU(2)$\times$SU(2) based gauge-theory model, describing wave- and particle-like 
aspects of electromagnetic disturbances within the so far experimentally/observationally investigated spectrum.     
\end{abstract} 

\end{titlepage}

\section{Introduction}

Boltzmann's statistical approach to kinetic gas theory can be considered an anticipation of 
Quantum Physics. Assuming for simplicity a single atomic species of mass $m$, his equation reads
\eqb
\label{BoltzmannEq}
\left(\pd_t+\dot{x}_i\pd_{x_i}+\frac{1}{m} F_i\pd_{\dot{x}_i}\right)f(\vec{x},\dot{\vec{x}},t)=\left.\pd_t f\right|_{\tiny\mbox{coll}}\ \ \ \ (i=1,2,3)\,,
\eqe
where $\vec{F}$ is an external force field, and the term on the right-hand side denotes the collision 
integral -- a functional of the phase-space probability distribution $f$. Eq.\,(\ref{BoltzmannEq}) would describe a time-reversal invariant evolution of $f$ like in Classical Mechanics\footnote{That is, if $f(\vec{x},\dot{\vec{x}},t)$ 
satisfies Eq.\,(\ref{BoltzmannEq}) then so does $f(\vec{x},-\dot{\vec{x}},-t)$.} 
if the collision integral was identically zero or determined from Classical Mechanics itself. 
However, the collision integral's intrinsic indeterminism, expressed through probabilistic changes from initial to final scattering states ({\sl molecular chaos}), selfconsistently underlies the concept of the 
probability distribution $f$ and its time-reversal non-invariant evolution. Presently, we use 
Quantum Mechanical or Quantum Field Theoretical amplitudes to compute $\left.\pd_t f\right|_{\tiny\mbox{coll}}$ from 
first principles for dilute gases (typical scattering lengths smaller than mean interparticle distance). 
The arrow of time, expressing the asymptotic attainment of an ergodic (thermal) equilibrium state of maximum entropy 
as a consequence of $f$'s evolution via Boltzmann's equation (\ref{BoltzmannEq}), thus is a direct consequence 
of the indeterminism inherent to the collision integral, and our modern understanding of {\sl molecular chaos} is that this integral be expanded into positive powers of $\hbar$ -- Planck's (reduced) quantum of 
action\footnote{In the formal limit $\hbar\to 0$ the quantity $\left.\pd_t f\right|_{\tiny\mbox{coll}}$ is given by Classical Mechanics or vanishes.}. The purpose of the present article is to discuss the emergence 
of Quantum Thermodynamics 
in pure SU(2) Yang-Mills theories and to explore some of its consequences which seem to extend 
beyond thermodynamics. 

In contrast to thermalization of a dilute gas of 
massive (bosonic) particles, such as atoms or molecules, by virtue of 
the collision integral in Boltzmann's equation the emergence of the Bose-Einstein quantum distribution is a much 
more fundamental and direct affair in pure Yang-Mills theory. This is because classical SU(2) 
Yang-Mills theory on the Euclidean spacetime ${\bf R}^4$ 
provides for (anti)selfdual and temporally periodic gauge-field configurations -- so-called Harrington-Shepard (HS) (anti)calorons\footnote{HS (anti)calorons represent the basic constituents of the thermal ground state in the deconfining phase of SU(2) Yang-Mills Quantum Thermodynamics \cite{HH2004,Hofmann2005,HofmannBook}.} of topological 
charge modulus $|k|=1$ and trivial holonomy \cite{HS1977,Nahm,KraanVanBaal,LeeLu,Diakonov,'tHooft,RebbiJackiw} -- 
which exhibit boundary behavior around a central spacetime point $x_0$ defining $\hbar$. Indeterminism of {\sl Minkowskian} processes involving (anti)caloron centers is an immediate consequence 
which, together with spatial independence, implies the Bose-Einstein distribution 
for thermal photons. Interactions of photons with massive vector 
modes (adjoint Higgs mechanism) are mediated by effective 
vertices, which occur through (anti)caloron centers in the vicinity of the 
center liberating a photon, are feeble, and, 
for sufficiently high temperatures, amount to a slight rescaling of 
$T\to T^\prime$ in thermodynamical quantities 
\cite{SchwarzGiacosaHofmann2007}. The transition from 
$T\to T^\prime$ can be interpreted as a 
re-thermalization due to a collision 
integral in the sense of Eq.\,(\ref{BoltzmannEq}), 
associated with loop integration in the photon's polarization 
tensor \cite{SchwarzGiacosaHofmann2007,LudescherHofmann2008}.          

The very notion of a Minkowskian spacetime and the Poincar\'e group, however, relates to the static 
structure of HS (anti)calorons spatially far 
from point $x_0$ \cite{GrandouHofmann2015} if at least two gauge-group 
factors SU(2) of disparate Yang-Mills scales \cite{Hofmann2015WP} are invoked. 
More specifically, we will argue that the spatial peripheries of HS (anti)calorons enable the propagation 
of coherent, wave-like disturbances, as introduced, e.g., by classically oscillating electric charges, through 
re-polarizations of electric and magnetic dipole densities. The wave-particle duality of 
electromagnetic disturbances would thus be understood in terms of the spatial peripheries and centers of 
(anti)selfdual, Euclidean field 
configurations in SU(2) Yang-Mills theory. 

This paper is organized as follows. In Sec.\,\ref{DTGS} the 
construction of the thermal ground state for deconfining SU(2) Yang-Mills 
thermodynamics, the properties of its effective thermal excitations, and the physics of 
effective radiative corrections are sketched for the reader's convenience. Sec.\,\ref{thermalPhotons} 
reviews observational and theoretical reasons for the postulate 
that an SU(2) rather than a U(1) gauge principle underlies the 
fundamental description of thermal photon gases \cite{Hofmann2005}. A discussion of how 
the spatial periphery of HS (anti)calorons provides electric and magnetic 
dipole densities, which (i) are protected against (anti)caloron overlap, 
are (ii) associated with the electric and magnetic polarizability of the ground state as induced by external 
disturbances $\vec{E}$ and $\vec{B}$, and (iii) propagate these disturbances at a finite speed, 
being independent of frequency and intensity within certain 
bounds governed by the Yang-Mills scale $\Lambda$, is 
performed in Sec.\,\ref{emwaves}. Both, propagation of isolated 
monochromatic electromagnetic wave and of waves in a thermal ensemble are addressed and confronted 
with experiment. Sec.\,\ref{photons} elucidates how short wave lengths, 
which probe (anti)caloron centers, provoke indeterministic reponses. Quantities, which associate with 
(anti)caloron centers, are interpretable 
in a Minkowskian spacetime -- a concept induced by 
overlapping (anti)caloron peripheries -- if they do not depend on 
analytic continuation from imaginary to real time. Only integral, gauge-invariant 
quantities qualify. In particular, the integral of the Chern-Simons current $K_\mu$ over the 3-sphere of vanishing 
4D radius (topological charge) and therefore the (anti)caloron action 
$S_{\rm C,A}=8\pi^2/e^2$ is physical in this sense. Here $e$ denotes the effective gauge coupling. It turns out 
that $S_{\rm C,A}=\hbar$ \cite{Kaviani2012,Krasowski2014}. As it seems, $\hbar$ is the only physical 
quantity which can be associated with the center of an (anti)caloron. 
What can be measured in response of probing such a center is $\hbar$ in combination with classical 
physical quantities such as frequency or wave number of a disturbance which, temporarily, is 
propagated by spatial (anti)caloron peripheries. We show that the 
statistical independence of the emission of (monochromatic) quanta of energy and 
momenta implies the Bose-Einstein distribution. Sec.\,\ref{SC} summarizes the present work.     

If not stated otherwise we work in (super-)natural units $c=\hbar=k_B=1$ from now on, $c$ denoting the speed of light in 
vacuum, and $k_B$ is Boltzmann's constant.  

\section{Mini-Review on deconfining thermal ground state, excitations, and radiative corrections\label{DTGS}}

Let us briefly review how the thermal ground state in the deconfining phase of thermal SU(2) Quantum Yang--Mills 
theory emerges out of HS (anti)calorons of topological charge modulus $|k|=1$ \cite{HS1977}. 
A crucial observation is that the energy momentum tensor $\theta_{\mu\nu}$ vanishes 
identically on these (anti)selfdual, periodic Euclidean field configurations in isolation. This implies that HS (anti)calorons do not propagate. 
Moreover, their spatial peripheries are static, meaning that any adiabatically slow  
process that induces a finite density of centers does not generate any propagating disturbance on 
distances larger than the spatial radius $R$ attached to an (anti)caloron center. 
On the other hand, given that the Euclidean time dependence of 
field-strength correlations within the central region set by $R$ 
spatially can be coarse grained into a mere choice of gauge for the 
inert, adjoint scalar field $\phi$ of space-time independent modulus $|\phi|$ \cite{HH2004,Hofmann2005} and taking into account that there is a preferred value $R\sim |\phi|^{-1}$ at a given temperature $T$ \cite{Kaviani2012} it is clear 
that no (potentially to be continued) gauge-invariant Euclidean time dependencies leak out from a center to  
its periphery. (Static peripheries cannot resolve and therefore deform centers. However, their spatial 
overlap, as facilitated by dense packing of centers, introduces a departure from (anti)selfduality and thus finite energy density and pressure \cite{Hofmann2005}.) Moreover, topologically trivial, propagating 
disturbances are governed by an action which is of the same form as the 
fundamental Yang-Mills action since their (Minkowskian) time 
dependence can be introduced adiabatically into the physics of 
overlapping (anti)caloron peripheries and since 
off-shellness, introduced by just-not-resolved and 
thus integrated edges of (anti)caloron centers, do not change the form of the action thanks 
to perturbative renormalizability, see below.          

Let us be more specific. For gauge group SU(2) the 
Harrington-Shepard (HS) caloron (C) -- a gauge-field configuration whose spacetime 
components $A_\mu$ ($\mu=4,1,2,3$) assume values in the SU(2) Lie algebra su(2) -- 
is given as (antihermitian group generators $t_a$ ($a=1,2,3$) with 
tr\,$t_at_b=-\frac12\delta_{ab}$)
\eqb
\label{HScaloron}
A_\mu=\bar{\eta}_{\mu\nu}^a t_a \pd_\nu\log\Pi(\tau,r)\,,
\eqe
where $r\equiv|\vec{x}|$, $\bar{\eta}_{\mu\nu}^a$ denotes the antiselfdual 
't Hooft symbol \cite{'tHooft}, $\bar{\eta}_{\mu\nu}^a=\epsilon^a_{\mu\nu}-\delta_{a\mu}\delta_{\nu 4}+
\delta_{a\nu}\delta_{\mu 4}$ ($\epsilon^a_{\mu\nu}$ the totally antisymmetric symbol in three dimension 
with $\epsilon^1_{23}=1$ and $\epsilon^a_{\mu\nu}=0$ for $\mu=4$ or $\nu=4$). The prepotential $\Pi(\tau,r)$ with 
\eqb
\label{HSprepotential}
\Pi(\tau,r)=1+\frac{\pi\rho^2}{\beta r}\frac{\sinh\left(\frac{2\pi r}{\beta}\right)}{\cosh\left(\frac{2\pi r}{\beta}\right)-\cos\left(\frac{2\pi\tau}{\beta}\right)} 
\eqe 
is derived by an infinite superposition of the temporally shifted prepotential $\Pi_0(x)$ of 
a singular-gauge instanton \cite{'tHooft,RebbiJackiw} with topological charge $k=1$ on ${\bf R}^4$ to render $\Pi(\tau,r)$ periodic in $\tau$. One has 
\eqb
\label{R4instperpot}
\Pi_0(x)=1+\frac{\rho^2}{x^2}\,,
\eqe
where $\rho$ is the instanton scale parameter. The associated antiselfdual field configuration (A) is obtained in replacing $\bar{\eta}_{\mu\nu}^a$ by $\eta_{\mu\nu}^a$ (selfdual 't Hooft symbol) 
in Eq.\,(\ref{HScaloron}). Configuration (\ref{HScaloron}) is singular at $\tau=r=0$ where the topological charge $k=1$ on $S_1\times{\bf R}^4$ is localized in the sense that the integral of the 
Chern-Simons current $K_\mu=\frac{1}{16\pi^2}\epsilon_
{\mu\alpha\beta\gamma}\left(A^a_\alpha\pd_\beta A^a_\gamma+
\frac13\epsilon^{abc} A_\alpha^a A_\beta^b A_\gamma^c \right)$ over a three-sphere $S^3_\delta$ of 
radius $\delta$, which is centered at $\tau=r=0$, is unity for $\delta\to 0$. Since the configuration C of Eq.\,(\ref{HScaloron}) is selfdual (and the associated configuration A is antiselfdual) the action of the 
HS (anti)caloron is given in terms of its topological charge $k=\pm 1$ and the gauge-coupling constant $g$ as
\eqb
\label{actHS}
S_C=S_A=\frac{8\pi^2|k|}{g^2}=\frac{8\pi^2}{g^2}\int_{S^3_\delta} d\Sigma_\mu K_\mu=\frac{8\pi^2}{g^2}\,.
\eqe
Since Eq.\,(\ref{actHS}) holds 
in the limit $\delta\to 0$ the action $S_C=S_A$ as a 
Minkowskian intepretation (trivial analytic continuation). Based on \cite{loopExp} and on the fact 
that the thermal ground state emerges from $|k|=1$ caloron/anticalorons, whose scale parameter $\rho$ essentially 
coincides with the inverse of maximal resolution, $|\phi|^{-1}$, in 
deconfining SU(2) Yang-Mills thermodynamics, it was argued 
in \cite{Kaviani2012}, see also \cite{Krasowski2014}, that $S_C$ and $S_A$ both equal 
$\hbar$ if the effective theory emergent from a spatial coarse-graining, see below, is to be interpreted as a 
local quantum field theory. With \cite{GPY1983}, see also \cite{GrandouHofmann2015}, 
we now investigate how the field strengths of C and A look like away from their centers at $\tau=r=0$. 

For $|x|\ll \beta$ one has
\eqb
\label{prepotentialx<<beta}
\Pi(x)=(1+\frac{\pi}{3}\frac{s}{\beta})+\frac{\rho^2}{x^2}+O(x^2/\beta^2)\,,
\eqe
where $s$ is given as
\eqb
\label{sdef}
s\equiv \pi\frac{\rho^2}{\beta}\,.
\eqe
From Eqs.\,(\ref{prepotentialx<<beta}) and 
(\ref{HScaloron}) one obtains for $|x|\ll \beta$ the following expression for $F_{\mu\nu}=\frac12\epsilon_{\mu\nu\kappa\lambda}F_{\kappa\lambda}\equiv\tilde{F}_{\mu\nu}$ on C
\eqb
\label{Finside}
F_{\mu\nu}^a=-4{\rho^\prime}^2 \frac{\bar{\eta}^a_{\alpha\beta}}{(x^2+{\rho^\prime}^2)^2} 
I_{\alpha\mu}I_{\beta\nu}+O(x^2/\beta^4)\,,
\eqe
where $I_{\alpha\mu}\equiv\delta_{\alpha\mu}-2\frac{x_\alpha x_\mu}{x^2}$. 
At small four-dimensional distances $|x|$ from the caloron center the field strength 
thus behaves like the one of a singular-gauge instanton with a 
renormalized scale parameter ${\rho^\prime}^2=\frac{\rho^2}{1+\frac{\pi}{3}\frac{s}{\beta}}$. For $|x|\ll \beta$ the field strength tensor $F_{\mu\nu}$ thus exhibits a dependence on $\tau$ which would give rise to a nontrivial analytic 
continuation having no Minkowskian interpretation. 
For $r\gg\beta$ the selfdual electric and magnetic fields $E_i^a$ and $B_i^a$ are {\sl static}: 
\eqb
\label{EBlarge}
E^a_i=B_i^a\sim-\frac{\frac{\hat{x}^a\hat{x}_i}{r^2}-\frac{1}{rs}(\delta^a_i-3\hat{x}^a\hat{x}_i)}
{(1+\frac{r}{s})^2}\,.
\eqe
Here $\hat{x}_i\equiv\frac{x_i}{r}$ and $\hat{x}^a\equiv\frac{x^a}{r}$. 
A simplification of Eq.\,(\ref{EBlarge}) occurs for $\beta\ll r\ll s$ as
\eqb
\label{EBlargeDyon}
E^a_i=B_i^a\sim-\frac{\hat{x}^a\hat{x}_i}{r^2}\,. 
\eqe
This is the field of a static non-Abelian monopole of unit electric and magnetic charges (dyon). 
For $r\gg s\gg\beta$ Eq.\,(\ref{EBlarge}) reduces to 
\eqb
\label{EBlargeDipole}
E^a_i=B_i^a\sim s\,\frac{\delta^a_i-3\,\hat{x}^a\hat{x}_i}{r^3}\,,
\eqe  
representing the field strength of a static, selfdual non-Abelian dipole field. The 
dipole moment $p_i^a$ of the latter is given as
\eqb
\label{dipolemom}
p_i^a=s\,\delta_i^a\,.
\eqe
For A one simply replaces $E^a_i=B_i^a$ by $E^a_i=-B_i^a$ 
in Eqs.\,(\ref{EBlarge}), (\ref{EBlargeDyon}), and (\ref{EBlargeDipole}). 

It is instructive to discuss a slight deformation of the HS caloron towards non-trivial 
holonomy, keeping $s$ fixed and maintaining selfduality \cite{LeeLu}. A holonomy 
$u\ll \pi/\beta$ then produces a nearly massless and de-localized 
magnetic monopole (charged w.r.t. U(1)$\subset$SU(2) left unbroken 
under $A_4(|\vec{x}|\to\infty)=u t^3\not=0$ where $t^a$ ($a=1,2,3$) is the hermitian generator of SU(2),  
normalized to tr\,$t^at^b=\frac12\,\delta^{ab}$) and its localized massive antimonopole, the latter centered at 
$r\sim 0$ -- a position which nearly coincides with the spatial 
locus of topological charge of the (anti)caloron \cite{LeeLu}. 
The centers of the mass densities of both particles are separated by $s$. 
For $r\ll s$ the massive antimonopole appears like a purely magnetic charge. 
However, as $r$ increases beyond $s$ this magnetic charge is increasingly 
screened by the presence of the delocalized magnetic monopole such that 
the non-Abelian, selfdual field strength of 
Eq.\,(\ref{EBlargeDipole}) prevails (no reference to the 
scale $u$)\footnote{In contrast, the definition of an Abelian field strength, 
see \cite{'tHooftMonop}, requires that $u\not=0$ to be able to 
define the su(2) unit vector $\hat{A}_4$ everywhere except for the two central points of the 
magnetic charge distributions.}. Moreover, it was shown in \cite{Diakonov} that 
$u\ll \pi/\beta$ leads to monopole-antimonopole attraction under the 
influence of small field fluctuations. This renders the interpretation of 
$s$ as the scale of magnetic monopole-antimonopole separation irrelevant for the physics 
of (slightly deformed) HS (anti)caloron peripheries.   

Let us now come back to the question how field $\phi$ emerges thanks to HS (anti)calorons. 
We have discussed in \cite{HofmannBook} why the following definition 
of a family of phases associated with the inert field $\phi$ is 
unique and, as a whole, transforms homogeneously under fundamental gauge transformations   
\begin{eqnarray}
  \{\hat{\phi}^a\}&\equiv&\sum_{C,A}{\rm tr}\!\!\int d^3x \int d\rho \,t^a\,
F_{\mu\nu} (\tau,\vec 0) \, \{ (\tau,\vec 0),(\tau,{\vec x})\} \nonumber\\[3pt]
&&\times\, F_{\mu\nu} (\tau,{\vec x}) \{(\tau,{\vec x}),(\tau,{\vec 0})\} ,
\label{definition}
\end{eqnarray}
where the Wilson line $\{(\tau,\vec 0),(\tau,\vec x)\}$ is defined as
\begin{equation}
\label{abk}
\begin{split}
\{(\tau,\vec 0),(\tau,\vec x)\}&\equiv
{\cal P} \exp \left[ i \int_{(\tau,\vec 0)}^{(\tau,\vec x)} dz_{\mu} \, A_{\mu}(z) \right],  \\
\{(\tau,\vec x),(\tau,\vec 0)\}
&\equiv \{(\tau,\vec 0),(\tau,\vec x)\}^\dagger\,,
\end{split}
\end{equation}
the integration in Eq.\,(\ref{abk}) is along the straight spatial 
line connecting the points $(\tau,\vec 0)$ and $(\tau,\vec x)$, and 
the sum is over configuration C of Eq.\,(\ref{HScaloron}) and 
its antiselfdual partner A. Moreover, $F_{\mu\nu}\equiv \pd_\mu A_\nu-\pd_\nu A_\mu-[A_\mu,A_\nu]$ denotes the Yang-Mills field-strength tensor, and the symbol ${\cal P}$ demands path ordering. On C and A path ordering actually is
obsolete since the spatial components of the gauge field represent a hedge-hog 
configuration which fixes the direction in su(2) in terms of 
the direction in 3-space. As a consequence, all infinitesimal contributions to the line integral in 
Eq.\,(\ref{abk}) commute, and in summing them their order is irrelevant. One can show 
\cite{HH2004,HofmannBook} that in performing the integrations over 
$\vec{x}$ and $\rho$ in Eq.\,(\ref{definition}) and by re-instating 
temporal shifts $\tau\to \tau+\tau_{C,A}$, the family $\{\hat{\phi}^a\}$ is parameterized, modulo global gauge rotations, 
by four real parameters, two for each ``polarization state'' for harmonic motion 
in a plane of su(2). This uniquely associates a linear differential operator ${\cal D}$ of order 
two with $\{\hat{\phi}^a\}$: ${\cal D}\equiv\partial^2_\tau+\left(\frac{2\pi}{\beta}\right)^2$. 
Moreover, one shows that the result of the $\rho$-integration, which depends cubically 
on its upper cutoff $\rho_u$, hence is sharply dominated by $\rho_u$ whatever the 
value of this cutoff turns out to be. 

Operator ${\cal D}$ exhibits an explicit temperature ($\beta$) dependence. However, due to the fact 
that the action in Eq.\,(\ref{actHS}), which determines the 
weight in the partition function that is introduced by HS (anti)calorons, 
is not temperature dependent such an explicit temperature dependence must not appear in 
the effective, thermal Yang-Mills action (ETYMA) obtained from a spatial coarse-graining in combination 
with integrating out these (anti)calorons. Therefore, in deriving the part of 
ETYMA, which is solely due to the field $\phi$, by demanding it to be stationary 
w.r.t. variations in $\phi$ at a fixed value of $\beta$ 
(Euler--Lagrange equation) the explicit $\beta$ dependence in 
${\cal D}$ is to be absorbed into the $\phi$-derivative of a 
potential $V(\phi)$. Demanding consistency of a first-order Bogomol'nyi-Prasad-Sommerfield (BPS) 
equation, which needs to be satisfied by $\phi$ owing to the fact that it embodies 
spatial field-strength correlations on (anti)selfdual gauge-field configurations, one derives 
the following first-order equation for $V$     
\eqb
\label{eomPot}
\frac{\partial V(|\phi|^2)}{\partial
  |\phi|^2}=-\frac{V(|\phi|^2)}{|\phi|^2}\,,
\eqe
whose solution reads
\begin{equation}
\label{solPot}
V(|\phi|^2)=\frac{\Lambda^6}{|\phi|^2}\,,
\end{equation}
where $\Lambda$ denotes an arbitrary mass scale (the Yang--Mills
scale). This implies that 
\eqb
\label{phismodulus}
|\phi|=\sqrt{\frac{\Lambda^3\beta}{2\pi}}\,.
\eqe
Scale $|\phi|^{-1}$ represents a minimal length scale in evaluating the consequences of 
ETYMA. Therefore, $\rho_u\sim |\phi|^{-1}$. The condition $s_u\gg\beta$, which is required for Eqs.\,(\ref{EBlargeDyon}) 
and (\ref{EBlargeDipole}) to actually represent static field strengths, is always satisfied provided that the dimensionless temperature 
$\lambda\equiv\frac{2\pi T}{\Lambda}\gg 1$. Namely, one then has
\eqb
\label{rhorequ}
\frac{s_u}{\beta}\equiv\pi\left(\frac{\rho_u}{\beta}\right)^2\sim 
\pi\left(\frac{\lambda^{3/2}}{2\pi}\right)^2=\frac{\lambda^{3}}{4\pi}\gg 1\,.
\eqe
Also, it is true that 
\eqb
\label{rhouvsbeta}
\frac{\rho_u}{\beta}\sim\frac{\lambda^{3/2}}{2\pi}\gg 1\,.
\eqe
That the condition $\lambda\gg 1$ is satisfied in the deconfining phase of SU(2) Yang-Mills thermodynamics is a 
consequence of the evolution equation for the effective coupling $e$. This evolution follows from 
the demand of thermal consistency of the Yang-Mills gas of 
non-interacting thermal quasi-particle fluctuations and 
their thermal ground state \cite{Hofmann2005}, based on ETYMA density 
\begin{equation}
\label{fullactden}
{\cal L}_{{\rm eff}}[a_\mu]={\rm tr}\,\left(\frac12\,
  G_{\mu\nu}G_{\mu\nu}+(D_\mu\phi)^2+\frac{\Lambda^6}{\phi^2}\right)\,.
\end{equation}
In Eq.\,(\ref{fullactden}) $G_{\mu\nu}=\partial_\mu a_\nu-\partial_\nu
a_\mu-ie[a_\mu,a_\nu]\equiv G^a_{\mu\nu}\,t_a$ denotes the field
strength of the {\it effective} trivial-topology gauge field $a_\mu=a_\mu^a\,t_a$,
$D_\mu\phi=\partial_\mu\phi-ie[a_\mu,\phi]$, and $e$ is the effective
gauge coupling. The latter takes the value $e=\sqrt{8}\pi$ almost everywhere in the deconfining 
phase (in natural units $\hbar=k_B=c=1$) \cite{DolanJackiw,GiacosaHofmann,HofmannBook}. One can show \cite{Kaviani2012,HofmannBook} that ${\cal L}_{{\rm eff}}$ is uniquely 
determined as in Eq.\,(\ref{fullactden}), resting on the facts that the 
effective $k=0$ field $a_\mu$ is governed by the first 
term due to perturbative renormalizability \cite{'tHooftVeltman}, gauge 
invariance fixes the second term, and no higher-dimensional mixed operators, involving 
fields $a_\mu$ and $\phi$, may appear due to the impossibility of the former to resolve the 
physics leading to the latter (inertness). The action density of 
Eq.\,(\ref{fullactden}) predicts the existence of one massless and two massive (adjoint Higgs mechanism, thermal quasi-particle excitations) directions in su(2) provided that their interactions are feeble and justifiedly expandable 
into a growing (likely finite, see below) number of vertices. In unitary-Coulomb gauge (a completely fixed, physical gauge) constraints on admissible four-momentum transfers can be stated precisely. 
These constraints imply a rapid numerical convergence of radiative 
corrections \cite{SchwarzGiacosaHofmann2007,KavianiHofmann}, and, by counting the number of constraints 
versus the number of radial loop variables in dependence of loop number, it was conjectured in \cite{Hofmann2006,HofmannBook} that one-particle irreducible bubble diagrams vanish, 
starting from a finite loop number. Note that Eq.\,(\ref{rhouvsbeta}) states the independence of $\phi$'s modulus on Euclidean time $\tau$, and Eqs.\,(\ref{rhorequ}) and (\ref{EBlargeDipole}) 
indicate that an (anti)selfdual static dipole field only 
emerges spatially far from the central region of an (anti)caloron, the latter being bounded by a spatial sphere 
of radius $|\phi|^{-1}\sim\rho_u$.

\section{Mini-Review on the postulate SU(2)$_{\rm CMB}$\\ (thermal photon gases)\label{thermalPhotons}}

In \cite{Hofmann2005} we have postulated that thermal photon gases, fundamentally seen, 
should be subject to an SU(2) rather than a U(1) gauge principle. 

Theoretically, such a postulate rests on the facts that in the deconfining phase of 
Yang-Mills thermodynamics the gauge symmetry SU(2) is broken to U(1) by 
the field $\phi$ and that the interaction between 
massive and massless excitations is feeble with the exception of the 
low-frequency regime at temperatures not far above the critical temperature $T_c$ 
for the deconfining-preconfining phase transition \cite{SchwarzGiacosaHofmann2007,LudescherHofmann2008,FalquezHofmannBaumbach2011}. 
Observationally, however, the physics of the deconfining-preconfining 
phase boundary \cite{Hofmann2009,HofmannBook}, the presence of a nontrivial thermal ground 
state, giving rise to massive quasi-particle fluctuations and therefore an equation 
of state $p=p(\rho)\not=\frac13\rho$, and 
feeble radiative effects influencing the propagation properties of the massless mode \cite{SzopaHofmann2007,Hofmann2013,HofmannBook} allow to confront the SU(2) 
postulate with reality. As for the former, a highly significant cosmological radio excess at 
frequencies $\nu\le 1\,$GHz \cite{Arcade2}, when considered in the SU(2) 
framework, links the evanescence of low-frequency electromagnetic waves belonging to the Cosmic Microwave Background (CMB) 
to an (incomplete) condensation phenomenon involving screened and ultralight electric charges\footnote{In units, where $\hbar$ is re-instated as a dimensionful quantity, one has $e=\sqrt{8}\pi/\sqrt{\hbar}$ almost everywhere in the deconfining phase. 
This and the fact that the thermal ground state is sharpy dominated by (anti)caloron radii 
$\rho_u\sim |\phi|^{-1}$, see Sec.\,(\ref{DTGS}), imply 
that the (anti)caloron action equals $\hbar$ \cite{Kaviani2012}. 
According to Eq.\,(\ref{actHS}) this quantum of action is localized at the inmost spacetime point in an 
(anti)caloron center. Moreover, the fact that the (unitless)  QED fine-structure constant $\alpha$ is given as $\alpha=\frac{Q^2}{4\pi\hbar}$, where $Q$ denotes the charge 
of the electron, implies an electric-magnetically dual interpretation of the U(1) charge content \cite{'tHooftMonop} of SU(2) field configurations \cite{Kaviani2012}: $Q\propto 1/e$.}.  This gives rise to a (partial) Meissner 
effect, and hence frequencies smaller than the implied Meissner mass $m_\gamma$ do not propagate 
but constitute an ensemble of evanescent waves\footnote{In Sec.\,(\ref{emwaves}) we show 
that these low frequencies, indeed, associate with classical waves.}. 
As a result, a re-shuffling of spectral power, creating a maximum at zero frequency, 
takes place at small CMB frequencies. Because $m_\gamma$ is (critically and thus rapidly) 
increasing when $T$ is decreased below $T_c$ \cite{Hofmann2005} the observation of a 
spectral-excess anomaly in the CMB at small frequencies implies that the present baseline temperature of the CMB, 
$T_0=2.725\,$K, practically coincides with $T_c$. This fixes the Yang-Mills scale $\Lambda$ of the 
theory by virtue of $\Lambda=\frac{2\pi}{13.87}T_c\sim 10^{-4}$\,eV \cite{Hofmann2009} which prompts 
the name SU(2)$_{\rm CMB}$. Based on the precise experimental match $T_c=T_0$ 
and on the availability of the (practically one-loop exact) equation of state $p=p(\rho)$ 
of deconfining SU(2) Yang--Mills thermodynamics \cite{Hofmann2005}, a prediction 
of the CMB redshift ($z$) -- temperature ($T$) relation 
is accomplished \cite{Hofmann2015} which exhibits strong violations of conformal 
behavior at $T\sim 2\,T_0$ where $z\sim 2.1$ (conventionally: $z\sim 1$). 
As a consequence, the discrepancy between the redshift $z_{\rm re}$ for instantaneous re-ionization of 
the intergalactic medium, as extracted with $z_{\rm re}\sim 11$ from the depletion of peaks in the CMB $TT$ angular power spectrum by appealing to the conventional, conformal $z$--$T$ relation $T/T_0=z+1$ \cite{Planck2013}, 
and as observed with $z_{\rm re}\sim 6$ by detection of the Gunn--Peterson trough 
for $z\ge z_{\rm re}$ in high-redshift quasar spectra \cite{Becker}, is resolved 
\cite{Hofmann2015}. Finally, with $T_c=T_0$ one predicts that the temperature dependence of {\sl radiatively} 
induced effects at low frequencies 
such as anomalies in blackbody spectra \cite{SchwarzGiacosaHofmann2007,LudescherHofmann2008,FalquezHofmannBaumbach2011} 
(spectral gap, extending from zero to about 17\,GHz at $T\sim 5.4\,$Kelvin) as well as the thermal excitation of longitudinally propagating magnetic-field modes \cite{FalquezHofmannBaumbach2012} (several, partially superluminal, low-frequency branches whose combined energy densities 
match the order of magnitude of the field strength ($\sim 10^{-8}\,$Gauss) squared of intergalactic magnetic fields 
extracted from small-angle CMB anisotropies \cite{Widrow2002}).

\section{Dipole densities:\\ (Anti)caloron peripheries and wave propagation\label{emwaves}}

In this section we discuss the vacuum properties of classical Electromagnetism -- electric permittivity $\epsilon_0$ and magnetic susceptibility $\mu_0$ -- and their possible relation to the thermal-ground state aspects caused by (anti)caloron peripheries, see also \cite{GrandouHofmann2015}. It will become clear 
that, in describing thermal photon gases, classical aspects of the thermal 
ground state of SU(2)$_{\rm CMB}$ are limited to very low frequencies. 

We have seen by virtue of Eqs.\,(\ref{rhouvsbeta}) and (\ref{rhorequ}) that a probe being 
sensitive to spatial distances $r$ from a given (anti)caloron center, which are much greater 
than the scale $s_u$ ($s_u$ itself being much greater than the coarse-graining scale $\rho_u\sim |\phi|^{-1}$), 
detects the static (anti)selfdual dipole field of Eq.\,(\ref{EBlargeDipole}). The 
electromagnetic field, which propagates through the deconfining 
thermal ground state in absence of any explicit electric charges, is considered a 
monochromatic plane wave of wave length $l\sim r$. Such a field associates with a density of (anti)selfdual 
dipoles, see Eq.\,(\ref{EBlargeDipole}). Because they are given by $p_i^a=s_u\delta^a_i$ 
their dipole moments align along the direction of the exciting electric or magnetic field both in 
space and in su(2). Note that at this stage the definition of what is to be viewed as an 
Abelian direction in su(2) is a global gauge convention such that {\sl all} spatial 
directions of the dipole moment $p_i^a$ are a priori thinkable. In a thermal 
situation and unitary gauge $\phi=2\,t^3\,|\phi|$ we would 
thus set $a=3$ which implies that $\vec{p}=s_u\hat{e}_3$.

Per spatial coarse-graining volume $V_{\tiny\mbox{cg}}$ of radius $|\phi|^{-1}=\rho=\sqrt{\frac{\Lambda^3}{2\pi T}}$ with 
\eqb
\label{cgvol}
V_{\tiny\mbox{cg}}=\frac43\pi|\phi|^{-3}\,,
\eqe
the center of a selfdual HS caloron or 
the center of an antiselfdual HS anticaloron \cite{HofmannBook} resides. Note the large hierachy between $s_u$ 
(the minimal spatial distance to the center of a (anti)caloron, which allows to identify 
the static, (anti)selfdual dipole) and the radius of the sphere $|\phi|^{-1}$ defining 
$V_{\tiny\mbox{cg}}$, 
\eqb
\label{ratiophim1s}
\frac{s_u}{|\phi|^{-1}}=\frac12\lambda^{3/2}=25.83\,\left(\frac{\lambda}{\lambda_c}\right)^{3/2}\,.
\eqe
If the exciting field is electric, $\vec{E}_e$, then it 
sees {\sl twice} the electric dipole $p_i^a$ (cancellation of magnetic dipole between caloron and anticaloron), if it is magnetic, $\vec{B}_e$, it sees {\sl twice} the magnetic dipole $p_i^a$ (cancellation of electric dipole between caloron and anticaloron, $\vec{E}=-\vec{B}\ \ \Leftrightarrow\ \ -\vec{E}=\vec{B}$). 
To be definite, let us discuss the 
electric case in detail, which is characterized by $\vec{E}_e$. 
The modulus of the according dipole density $\vec{D}_e||\vec{E}_e$ is given as  
\eqb
\label{dipoledensity}
|\vec{D}_e|=\frac{2s_u}{V_{\tiny\mbox{cg}}}=
\frac{3}{4\pi}\Lambda^2\lambda^{1/2}_c\left(\frac{\lambda}{\lambda_c}\right)^{1/2}\,.
\eqe 
In classical electromagnetism the relation between the fields $\vec{E}_e$ and $\vec{D}_e$ is 
\eqb
\label{permvac}
\vec{D}_e=\epsilon_0\vec{E}_e\,,
\eqe
where 
\eqb
\label{measuredeps}
\epsilon_0=5.52703\times 10^7\,\frac{Q}{\mbox{V\,m}}
\eqe 
is the electric 
permittivity of the vacuum, and $Q=1.602\times 10^{-19}\,$A\,s denotes the elementary unit of electric charge (electron charge), both quoted in SI units.

According to electromagnetism the energy density $\rho_{\rm EM}$ carried by 
an external electromagnetic wave with $|\vec{E}_e|=|\vec{B}_e|$ is
\eqb
\label{eneE}
\rho_{\rm EM}=\frac{1}{2}(\epsilon_0|\vec{E}_e|^2+\frac{1}{\mu_0}|\vec{B}_e|^2)
=\frac{1}{2}(\epsilon_0+\frac{1}{\mu_0})|\vec{E}_e|^2\,.
\eqe
In natural units we have $\epsilon_0\mu_0=1/c^2=1$, and therefore\footnote{To set $\epsilon_0\mu_0=1$ 
is a short cut. This would have come out if we had 
treated the magnetic case explicitly.} one has $\mu_0=1/\epsilon_0$. 
Thus
\eqb
\label{eneready}
\rho_{\rm EM}=\epsilon_0|\vec{E}_e|^2\,.
\eqe
The $\vec{E}_e$-field dependence of $\rho_{\rm EM}$ is converted into a 
fictitious temperature dependence by demanding that the temperature of the 
thermal ground state of SU(2)$_{\tiny\mbox{CMB}}$ adjusts itself 
such as to accomodate $\rho_{\rm EM}$ in terms of its ground-state 
energy density $\rho_{\rm gs}$ \cite{Hofmann2005},
\eqb
\label{TE}
\rho_{\rm EM}=\rho_{\rm gs}=4\pi\Lambda^3 T\ \ \ \Leftrightarrow\ \ \ \ |\vec{E}_e|=\Lambda^2\sqrt{2\frac{\lambda_c}{\epsilon_0}}\left(\frac{\lambda}{\lambda_c}\right)^{1/2}\,. 
\eqe
Eq.\,(\ref{TE}) generalises 
the thermal situation of ground-state energy density (see below), 
where ground-state thermalisation is induced by a thermal ensemble 
of excitations, to the case where the thermal ensemble is missing but 
the probe field induces a fictitious temperature and energy density 
to the ground state. Combining Eqs.\,(\ref{dipoledensity}), (\ref{permvac}), 
and (\ref{TE}), and introducing the ratio $\xi$ 
 between the non-Abelian monopole charge $Q^\prime$ in the dipole and the (Abelian) 
 electron charge\footnote{In natural units, the actual charge of the 
monopole constituents within the (anti)selfdual dipole is $1/g$ where $g$ is the 
undetermined fundamental gauge coupling. This is absorbed into $\xi$.} 
$Q$, we obtain
\eab
\label{eps0}
\epsilon_0[Q\mbox{(V\,m)}^{-1}]&=&
\frac{3}{\sqrt{32}\pi}\left(\frac{\Lambda[\mbox{m}^{-1}]}{\Lambda[\mbox{eV}]}\right)^{1/2}\xi Q\sqrt{\epsilon_0[Q\mbox{(V\,m)}^{-1}]}\ \ \ 
\Leftrightarrow \nonumber\\ 
\epsilon_0[Q\mbox{(V\,m)}^{-1}]&=&\frac{9}{32\pi^2}\frac{\Lambda[\mbox{m}^{-1}]}{\Lambda[\mbox{eV}]}\,
(\xi Q)^2\,.
\eae 
Notice that $\epsilon_0$ does not exhibit any temperature dependence and thus no dependence 
on the field strength $\vec{E}_e$. It is a universal constant. 
In particular, $\epsilon_0$ does {\sl not} relate to the state of fictitious 
ground-state thermalisation which would associate to the rest frame of a local heat bath. 
To produce the measured value for $\epsilon_0$ as in Eq.\,(\ref{measuredeps}) the ratio $\xi$ in 
Eq.\,(\ref{eps0}) is required to be
\eqb
\label{xi}
\xi\equiv\frac{Q^\prime}{Q}=19.56\,.
\eqe 
Thus, compared to the electron charge, the charge unit associated with a (anti)selfdual non-Abelian dipole, 
residing in the thermal ground state, is gigantic. 
The discussion of $\mu_0$ proceeds in close analogy to the case of $\epsilon_0$.
(It would be $\mu_0^{-1}$ defining the ratio between 
the modulus of the magnetic dipole density and the magnetic flux density $|\vec{B}|$.) 
Here, however, the comparison between non-Abelian magnetic charge and 
an elementary, magnetic, and Abelian charge is not facilitated since the latter does not 
exist in electrodynamics.

The consideration above, linking the density of (anti)selfdual static dipoles in 
the thermal ground state to an exciting field-strength modulus 
$|\vec{E}_e|$ via a {\sl fictitious} temperature $T$, which represents the energy density 
of the thermal ground state in terms of the classical field-energy density introduced by 
$|\vec{E}_e|$, has assumed isolated propagation of a monochromatic plane wave. 
How would the argument that $\epsilon_0$ (and $\mu_0$) does not depend on 
$|\vec{E}_e|$ (or via $|\vec{E}_e|$ on $T$) have to be modified if a 
thermodynamical equilibrium subject to a genuine {\sl thermodynamical} 
temperature $T$ prevails? The condition that wavelength $l$ must be 
substantially larger than $s_u$ amounts to 
\eqb
\label{violcmb}
l=\frac{2\pi}{xT}\gg s_u=\frac{2\pi^2 T^2}{\Lambda^3}\ \Leftrightarrow \ 
\ x\ll\frac{1}{\pi} \left(\frac{\Lambda}{T}\right)^3\,,
\eqe 
where $x\equiv\frac{2\pi\nu}{T}$, and $\nu$ denotes the frequency of the wave. 
In particular, for $T=T_c$ (\ref{violcmb}) states that 
\eqb
\label{violcmbpart}  
x\ll \frac{1}{\pi}\,\left(\frac{2\pi}{\lambda_c}\right)^3\sim 0.0296\ \ \ \ \ (\lambda_c=13.87)\,.
\eqe
Considering that the maximum of Planck's spectral energy density 
$u_{\rm Planck}=\frac{2}{\pi} T^3\frac{x^3}{\e^x-1}$ occurs at $x=2.82$ we 
conclude that wave-like propagation in a thermodynamical situation is restricted 
to the deep Rayleigh-Jeans regime where spectral energy density is (classically) given as 
\eqb
\label{RJfreq}
u_{\rm RJ}=\frac{2}{\pi} T^3 x^2=8\pi T\nu^2\,.
\eqe
To convert $u_{\rm RJ}$ into an energy density it needs to be multiplied by 
a (constant) band width $\Delta\nu$. Notice that both, $\rho_{\rm RJ}\equiv u_{\rm RJ}\Delta\nu=
8\pi T\nu^2\Delta\nu$ and the energy density of the thermal ground state $\rho_{\rm gs}\equiv 
4\pi T\Lambda^3$, compare with Eq.\,(\ref{TE}), depend linearly on $T$. 
Therefore, an average electric field-strength modulus $|\vec{E}_e|$ in the 
Rayleigh-Jeans regime, defined as 
\eqb
\label{TEtherm}
\rho_{\rm RJ}=8\pi T\nu^2\Delta\nu=\rho_{\rm EM}=\epsilon_0|\vec{E}_e|^2\ \ \ \ \Leftrightarrow\ \ \ \ 
|\vec{E}_e|=2\nu\sqrt{\frac{\lambda_c\Lambda\Delta\nu}{\epsilon_0}}\left(\frac{\lambda}{\lambda_c}\right)^{1/2}\,,
\eqe 
also yields temperature independence of $\epsilon_0$,
\eqb
\label{epsilon_0therm}
\epsilon_0\equiv\frac{|\vec{D}_e|}{|\vec{E}_e|}=\frac{9}{64\pi^2} \frac{\Lambda[\mbox{m}^{-1}]}{\Lambda[\mbox{eV}]}(\tilde{\xi} Q)^2\times \frac{\Lambda^3}{\nu^2\Delta\nu}\Big[1\Big]\,,
\eqe 
where [1] indicates that the preceding fraction is to be evaluated in 
natural units ($\hbar=k_B=c=1$) so that it is dimensionless. The 
charge of a monopole in the dipole is represented by $\tilde{\xi} Q$. This charge now is 
perceived by the {\sl ensemble} of waves with 
frequencies contained in the band $\Delta\nu$. Since $\epsilon_0$ should be a frequency 
independent quantity we need to demand that 
\eqb
\label{screeningdipole}
\tilde{\xi}^2=C\frac{\nu^2\Delta\nu}{\Lambda^3}\,,
\eqe
where $C=2\xi^2$, compare with Eq.\,(\ref{eps0}). We conclude 
that the charge of a monopole making up the dipole 
as perceived by the ensemble of waves with 
frequencies contained in the band $\Delta\nu$ is increasingly 
screened with decreasing frequency $\nu$.   

Finally, from the condition $l\gg s_u$ and Eq.\,(\ref{TE}) one obtains (natural units)
\eqb
\label{uncertainty}
|\vec{E}_e|^4\nu\ll 8\Lambda^9\,.
\eqe
Relation (\ref{uncertainty}) needs to be obeyed by any classically propagating, monochromatic 
electromagnetic wave. Its violation indicates that the propagation of electromagnetic field energy no longer 
is mediated by an adiabatic time-harmonic modulation of the polarization state of electric and magnetic 
dipole densities of the vacuum, as provided by overlapping (anti)caloron peripheries, 
but by the quantum physics of (anti)caloron centers. 
Setting $\Lambda=\Lambda_{\rm CMB}$, (\ref{uncertainty}) is a strong restriction on admissible 
frequencies at commonly occurring intensities in the propagation of 
electromagnetic waves. Such a restriction, however, is not supported by experience. 
In \cite{Hofmann2015WP} it was therefore proposed to add flexibility to the value of $\Lambda$ 
by postulating a product SU(2)$_{\rm CMB}\times$SU(2)$_e$ of gauge groups with $\Lambda_e\sim m_e\sim 0.5\,$MeV, see 
also \cite{Hofmann2005,HofmannBook}, subject to a mixing angle of the 
unbroken (diagonal) subgroups which is adjusted depending on whether or not this gauge 
dynamics plays out in a thermal or nonthermal situation or any intermediate thereof. (In the present Standard Model of 
particle physics such a mixing between the U(1) subalgebra of SU(2)$_W$ and U(1)$_Y$, the latter being regarded as a fundamental gauge symmetry, is subject to a {\sl fixed} value of the associated Weinberg angle.) According to (\ref{uncertainty}) the large value of $\Lambda_e$ allows for the propagation of electromagnetic waves  
throughout the entire experimentally accessed frequency spectrum at commonly experienced intensities. 
However, by virtue of Eq.\,(\ref{TE}) those intensities usually relate to (fictitious) 
temperatures that are much lower than $T_{c,e}\sim 2.21\,\Lambda_e$. As a consequence, the hierarchy between $s_u$ and $|\phi|^{-1}$, taking place for $\lambda_c\ge\lambda$, actually is inverted in physical wave propagation subject to SU(2)$_e$. 
That is, the center of an (anti)caloron would extend well beyond a typical wavelength, 
thus in principle introducing hard-to-grasp nonthermal quantum behavior. Still, 
since (\ref{uncertainty}) does not depend on the concept of a 
temperature anymore we may regard it as universally valid: it needs to be satisfied 
by any monochromatic, classically propagating electromagnetic wave.

\section{Bose-Einstein distribution: (Anti)caloron centers and \\ quanta of 
energy and momentum\label{photons}}

The derivation of the dipole density in Eq.\,(\ref{dipoledensity}) has appealed to the 
independence and inertness of (anti)caloron centers in ``sourcing'' their respective peripheries, the latter 
supporting static dipole fields. This is consistent since fields propagating by virtue 
of peripheries never probe centers. The fact that the thermal ground state actually is a spatial 
arrangement of densely packed (anti)caloron centers, implying profound spatial overlaps of (anti)caloron 
peripheries, is implemented by Eq.\,(\ref{TE}) which assigns a finite energy density to this ground state in terms of some  temperature $T$ which, in turn, is determined by the field-strength modulus $|\vec{E}_e|$ in the sense of an adiabatic deformation of the isotropic, thermal situation. (Anti)caloron centers are probed, however, 
if the wavelength $l$ of a propagating disturbance approaches the 
value $s_u$ -- a situation when dipole moments induced by 
time-harmonic monopole accelerations, see Eq.\,(\ref{EBlargeDyon}), 
yield inconsistencies \cite{GrandouHofmann2015}. This mirrors the fact that Maxwell's equations are 
void of magnetic sources (locality).        

As the wave length $l$ of a would-be propagating disturbance substantially 
falls below $s_u$ we need to consider the physics inherent to the central region 
of an (anti)caloron which is anything but classical, 
see Sec.\,(\ref{thermalPhotons}). Thus classical quantities wave length $l$ and frequency $\nu$ both cease  
to be applicable as physical concepts. On the other hand, the only trivially 
continuable and thus physical quantity associated with the central region 
of an (anti)caloron is the quantum of action $\hbar$. This can be used 
to transmute the no longer applicable 
classical concepts $l$ and $\nu$ into valid concepts $|\vec{p}|=\hbar 2\pi l^{-1}$ (momentum modulus) and $E=\hbar 2\pi\nu$ (energy). Thus it is the indeterministic emission of a quantum of momentum and 
energy (photon) that is expected as the response of an (anti)caloron center 
to disturbances whose classical propagation is excluded. Since, apart from small correlative effects, 
which are induced by effective Yang-Mills vertices and computable in the theory (\ref{fullactden}), see  \cite{SchwarzGiacosaHofmann2007,KavianiHofmann,LudescherHofmann2008,FalquezHofmannBaumbach2011,FalquezHofmannBaumbach2012}, (anti)caloron centers act spatio-temporally independently, and thus the derivation of the mean photon occupation 
number $\bar{n}$ proceeds as usual. Namely, the Boltzmann weight $p_n$ of an $n$-fold photon event, each photon possessing energy $E$ in the thermal ensemble, is the $n$th power of the Boltzmann weight $p_1$ of a single photon event
\eqb
\label{Boltzmann}
p_1(x)=\e^{-x}\ \ \ \Rightarrow \ \ \ p_n(x)=\e^{-nx}\ \ \ \ \left(x\equiv\frac{E}{k_BT}\right)\,.
\eqe  
Therefore, the partition function $Z(x)$ reads
\eqb
\label{ParttionFunction}
Z(x)=\sum_{n=0}^\infty p_n(x)=\frac{1}{1-\e^{-x}}\,.
\eqe
Finally, mean photon number $\bar{n}(x)$ is given as 
\eab
\label{Photonnumber}
\bar{n}(x)&=&\frac{1}{Z(x)}\sum_{n=0}^\infty n p_n(x)\nonumber\\ 
&=&-\frac{d\log Z(x)}{dx}=\frac{1}{\e^{x}-1}\equiv n_B(x)\,,
\eae
where $n_B(x)$ denotes the Bose-Einstein distribution function.

\section{Summary\label{SC}}

In this contribution we have given a sketchy overview on how the thermal ground state emerges 
in SU(2) Yang-Mills theory in terms of a spatial coarse-graining over the center of an 
electric-magnetically (anti)selfdual (anti)caloron gauge-field configuration 
of topological charge modulus unity and trivial holonomy \cite{HS1977}, giving rise to an 
inert scalar field $\phi$, and a pure-gauge solution $a_\mu^{\rm gs}$ 
of the effective Yang-Mills field equations, sourced by $\phi$. 
Details of this process can be studied in \cite{HH2004,HofmannBook}.  

After motivating the postulate by observational facts that thermal photon gases should be subject to an SU(2) 
rather than a U(1) gauge principle we have subsequently 
addressed the question of how the SU(2) thermal ground 
state remains a valid concept in supporting the propagation of electromagnetic waves. 
Namely, the electric permittivity $\epsilon_0$ 
and the magnetic permeability $\mu_0$ of the vacuum, which 
are parameters of classical electromagnetism, are related to their respective dipole densities 
emerging from the peripheries of (anti)calorons while their central regions, $r\le |\phi|^{-1}$,  
are densely packed spatially. Both, $\epsilon_0$ and $\mu_0$ turn out 
to be temperature independent, and this derivation can be performed for both the case of an isolated, 
monochromatic wave and spectral bands within the deep Rayleigh-Jeans regime in a given 
black-body spectrum.  

The last part of this paper dealt with the physics implied by the central regions of 
(anti)calorons. We have argued here that the classical concepts frequency and 
wave length necessarily convert into quanta of energy and momentum (photon) by 
virtue of the localization of the quantum of action at the singularity of the field configuration 
$A_\mu$ of an (anti)caloron. Since, modulo small correlative effects -- computable in terms of radiative corrections in the effective theory -- central regions in (anti)calorons are, in a Minkowskian sense, 
spatiotemporally independent one concludes that, thermodynamically, the Boltzmann weight of 
an $n$-photon event in the gas factorizes into Boltzmann weights of a 
single-photon event. This implies the Bose-Einstein distribution function.
\vspace{0.1cm}\\ 
\noindent{\textbf{Acknowledgments:}} We would like to acknowledge interesting 
and stimulating conversations with Thierry Grandou, Steffen Hahn, Hugo Reinhardt, and Andreas Wipf.\vspace{0.1cm}\\ 
\noindent{{\textbf{Author Contributions:}} The original work contained in Secs.\,\ref{emwaves} 
and \ref{photons} is by RH. Other contributions, which are reviewed and cited 
in this work, may have involved additional authors.






\end{document}